\documentclass[12pt,color,epsfig,subfigure]{article}

\usepackage[T1]{fontenc}
\usepackage{rotating}
\usepackage{graphics}
\usepackage{amsmath}
\usepackage{amssymb}
\usepackage{pifont}
\usepackage{varioref}
\usepackage{cite}

\usepackage[dvips]{color}

\def\bcent	{\begin{center}}
\def\ecent	{\end{center}}
\def\bleft	{\begin{flushleft}}
\def\eleft	{\end{flushleft}}
\def\btab	{\begin{tabular}}
\def\etab	{\end{tabular}}
\def\bright	{\begin{flushright}}
\def\eright	{\end{flushright}}
\def\bnum	{\begin{enumerate}}
\def\enum	{\end{enumerate}}
\def\bitem	{\begin{itemize}}
\def\eitem 	{\end{itemize}}
\def\beqn       {\begin{equation}}
\def\eeqn       {\end{equation}}
\def\beqna      {\begin{eqnarray}}
\def\eeqna      {\end{eqnarray}}
\def\rarw       {\rightarrow}

\def\stop	{\widetilde{t_{1}}}
\def\stst       {\stop\stop^{*}}
\def\tt         {t\bar{t}}
\def\bb         {b\bar{b}}
\def\ww         {W^{+}W^{-}}
\def\pt		{p_{T}}
\def\et		{E_{T}}
\def\etjet	{E_{T}^{jet}}
\def\met	{E\!\!\!/_{T}}
\def\rpv        {R\!\!\!\!/}
\def\rarw       {\rightarrow}
\def\lint       {\mathcal{L}_{int}}
\def\mstop      {m_{\stop}}

\def\bb         {b\bar{b}}

\def\gev        {\: \rm GeV} 
\def\tev        {\: \rm TeV} 

\def\fb         {\: \rm fb}

\def\etal       {{\em et al.}}

\def\PRD#1#2#3  {{Phys.~Rev.} D{\bf{#1}}, #2 (#3)}
\def\PRL#1#2#3  {{Phys.~Rev.~Lett.} {\bf{#1}}, #2 (#3)}
\def\PLB#1#2#3  {{Phys.~Lett.} B{\bf{#1}}, #2 (#3)}
\def\PREP#1#2#3 {{Phys. Rep.} {\bf #1} (#3) #2}
\def\JHEP#1#2#3 {{JHEP} {\bf #1}, #2 (#3)}
\newcommand{\ijmp}[3] {Int. J. Mod. Phys. {\bf A#1} (#3) #2}           %
\begin{document}
\bcent
{\Large Search for the lightest scalar top quark in R-parity violating decays 
        at the LHC}
 \vglue 0.4cm
Debajyoti Choudhury$^{1}$, Madhumita Datta$^{2}$ and Manas Maity$^{2}$ \\[1ex]
$^1${ Department of Physics and Astrophysics, University of Delhi,   Delhi 110 007, India}\\
$^2${ Department of Physics, Visva-Bharati, Santiniketan 731 235, India}

\ecent
\begin{abstract}
The scalar partner of the top quark (the stop) is relatively light in
many models of supersymmetry breaking. We study the production of
stops at the Large Hadron Collider (LHC) and their subsequent decays
through baryon-number violating couplings such that the final state
contains no leptons. A detailed analysis performed using detector
level observables demonstrate that stop masses upto $\sim 600 \gev$
may be explored at the LHC depending on the branching ratios for such
decays and the integrated luminosity available. 
Extended to other analogous scenarios, the analysis will,
generically, probe even larger masses.
\end{abstract}

The status of global symmetries in particle physics is a much debated
one. Of particular relevance are fortuitous symmetries such as baryon
number ($B$) and lepton number ($L$). Apparently a consequence of the
particular choice of the field assignments within the Standard Model
(SM), each is left intact to all orders within perturbation theory,
only for a combination to be broken by nonperturbative effects. While
guaranteeing the stability of the proton, this leaves no room for the
observed baryon asymmetry in the universe. As is well known, the
latter needs, apart from enhanced levels of $CP$ violation and a phase
transition (or, at least, a non-equilibriated state  of the universe),
additional sources of $B$--violation (or $L$--violation for
leptogenesis).

This has led to a sustained study of models of physics going beyond
the SM admitting $B$-violation. Particularly well-motivated scenarios
pertain to grand unification and supersymmetry (SUSY), whether
considered separately or in conjunction. Indeed, low energy
supersymmetry is considered to be, perhaps, the most attractive
extension of the SM.  The most general superpotential of the minimal
supersymmetric standard model (MSSM) contains, apart from the
generalizations of the Yukawa terms and the Higgs potential of the SM,
additional terms violating $B$ and $L$.  Since, this would, nominally,
lead to rapid proton decay, such terms need to be eliminated, and a
global symmetry was introduced~\cite{salam_strathdee, fayet} for this
very purpose. Subsequently, a discrete symmetry, $R$-parity, was shown
to prohibit all dimension-four $B$-- or $L$--violating
terms~\cite{rpar}. However, as has 
long been
realised, proton decay
can be eliminated even when $R$-parity is broken 
provided
at least
one of $B$ and $L$ are conserved. This is a particularly fascinating
situation, for it admits a whole host of phenomenological consequences
absent in the $R$-parity conserving MSSM. Indeed, the collider
signatures of supersymmetric particle production now undergoes a
sea-change, for the lightest of the 
$SM$--superpartners
is no longer
stable. And while it has been argued that the introduction of
$R$-parity violation would deprive the MSSM of one of its most
atractive features, namely a natural Dark Matter candidate, note that
the gravitino could yet be stable on cosmological scales.

In this paper, we shall concentrate on the case where $B$ is broken,
but $L$ is not. Such a situation is well motivated in a class of
supersymmetric grand unified theories (GUTs)~\cite{Bviol}. 
This, then, allows for terms in the superpotential of the form
\beqna
W_{\rpv} & \ni & \lambda_{ijk}^{''}\bar{U}^{i}_{R}\bar{D}^{j}_{R}\bar{D}^{k}_{R} \ ,
\eeqna
where $\bar{U}^{i}_{R}$ and $\bar{D}^{j}_{R}$ denote the right-handed
up-quark and down-quark superfields respectively and the couplings
$\lambda_{ijk}^{''}$ are antisymmetric under the exchange of the last
two indices.  The corresponding Lagrangian can then be written in
terms of the component fields as
\beqn 
{\mathcal L}_{R\!\!\!/} = \lambda''_{ijk}
\left(u^c_i d^c_j \tilde{d}^*_k + u^c_i \tilde{d}^*_j d^c_k + \tilde{u}^*_i
  d^c_j d^c_k\right) + {\rm h.c.},
\label{lagrp}
\eeqn
thus allowing a squark to decay into a pair of anti-quarks\footnote{
  Unless otherwise mentioned, for every process, the charge conjugated
  process is also implied. }  violating {\em B} and {\em R}.

At a hadron collider, the couplings of eqn.(\ref{lagrp}) would lead to
resonant production of a scalar, with the rates being potentially
large if at least one of the quark fields belongs to the first
generation. This could, in principle, lead to spectacular signatures
both for dijets (invariant mass spectrum) or more complicated final
states~\cite{CDF_dijet, berger_harris_sullivan}. However, many of the
$\lambda''$ couplings are severely constrained by a variety of
low-energy observables~\cite{sher_probir_biswa,bcs_debrupa,reviews},
especially for operators involving at least two fields of the first
generation\footnote{Some of the couplings involving the third
  generation can be quite large though.}.  Given these two contrasting
pulls, we restrict ourselves to the ($\lambda''$--independent) pure
supersymmetric-QCD processes as far as squark production is concerned,
and consider a role for $\lambda''$ only in their decays.

The mechanism of SUSY breaking is open to speculation. However, most
such mechanisms postulate that the sfermion masses be unified at some
high scale, the extent of unification being model-dependent. This has
the additional benefit of suppressing potentially large flavour
changing neutral currents.  In coming down to the electroweak scale,
though, the masses would suffer substantial renormalization group
evoulution. Furthermore, the large Yukawa coupling of the top quark
plays a role not only in this evolution, but also in engendering a
mixing between the two super-partners of the top quark (the
stops). Together, these effects 
very often 
result in the lighter stop
being rendered the lightest of the strongly interacting superpartners.

Thus, it is phenomenologically interesting to search for top squarks
in the hadron colliders in the following scenario: (almost model
independent) pair production of $\stst$ via strong interaction
and decays thereafter via $R$-parity violating process controlled by
the parameter $\lambda^{''}$ \cite{dc_md_mm}.

It may be noted that there are scenarios beyond the SM other than SUSY
in which a strongly interacting elementary particle may decay into a
pair of quarks. GUTs, for example, abound in these~\cite{diquarks}.
Such ``diquarks'', of which squarks in the $R$-parity violating theory
constitute but one example, can occur in various forms. They could be
Lorentz scalars or vectors, $SU(2)_L$ singlets, doublets or triplets
and, under, $SU(3)_c$ transform either as a {\boldmath$ \bar 3$} or a
{\boldmath$6$}.  While only some of the above combinations could
appear in a trilinear coupling with two quark fields, others can
participate if a quadrilinear term involving, say the Higgs, is
considered\cite{Han:2010rf}. In the latter case, although the
effective diquark-quark-quark vertex would exist only after the
electrweak symmetry breaks, there is no {\it a priori} reason to relate the 
coupling to the quark mass terms. Diquarks, in particular, have been 
shown~\cite{Tait} to be leading contenders for explaining the 
anomalously large forward-backward asymmetry in top production as 
reported by CDF~\cite{CDF_afb}.

Yet another class of strongly interacting particles may decay into a
pair of jets\footnote{Distinguishing decays into a quark-antiquark
pair or into a quark-gluon pair from a decay into a quark-pair (or an
antiquark pair) is possible only if both the daughters carry a heavy
flavour.}, namely color-octets like axigluons~\cite{axigluon},
colorons~\cite{coloron} or Kaluza-Klein excitations of
gluons~\cite{KK-gluons} or even color-triplets like excited
quarks~\cite{excited}.  Such objects, though, can be formed as
resonances, often with unsuppressed couplings, rendering the search
for them to be much simpler~\cite{resonance_expt,
resonance_theory}. We will not consider such possibilities and would
limit ourselves to the more difficult task of diquarks (squarks) with
a small coupling to a pair of quarks.

\begin{figure}[tbp]
\vspace*{-5ex}
\includegraphics[width=0.5\textwidth, height=0.6\textwidth]{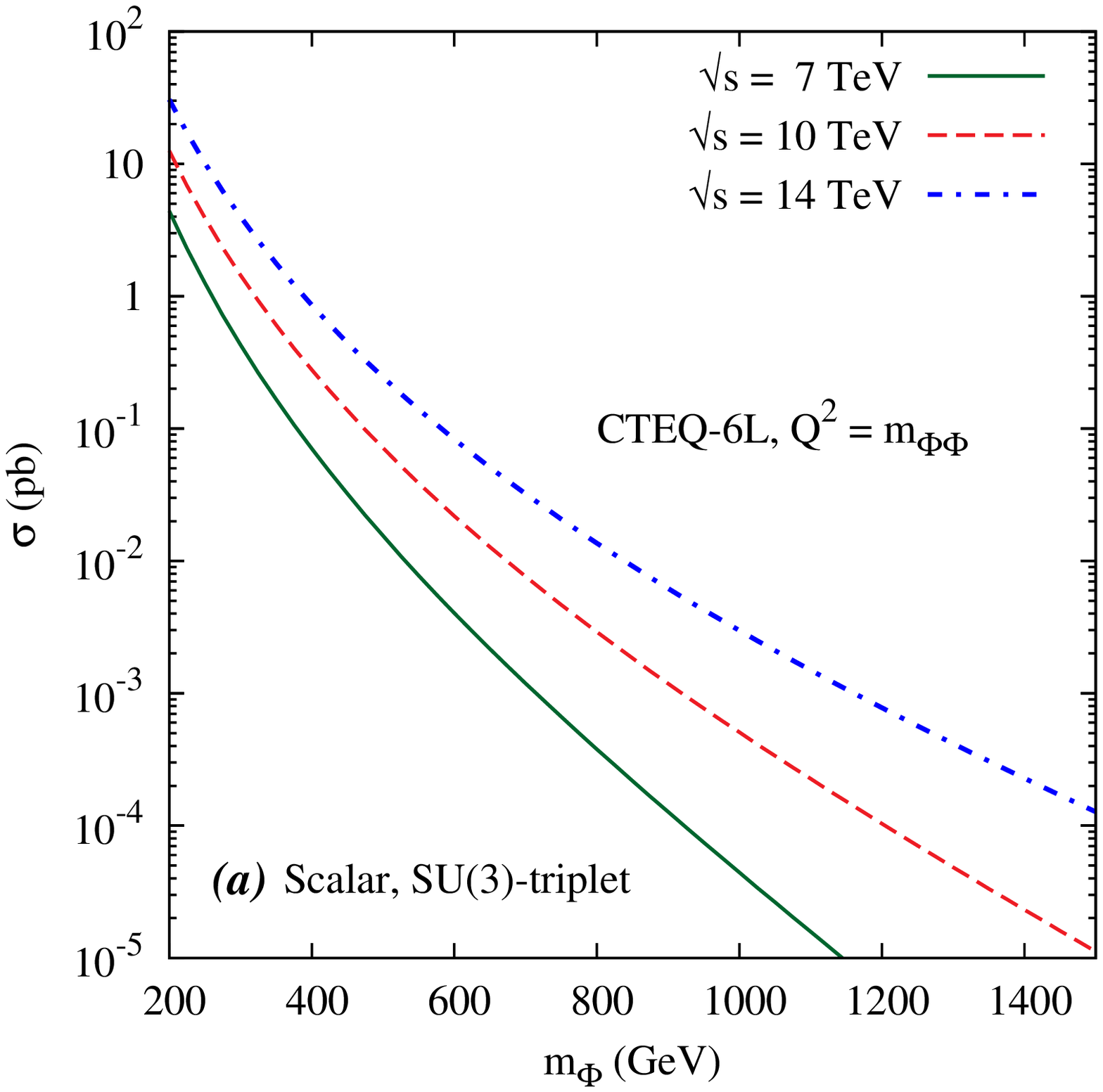}
\hfill
\includegraphics[width=0.5\textwidth, height=0.6\textwidth]{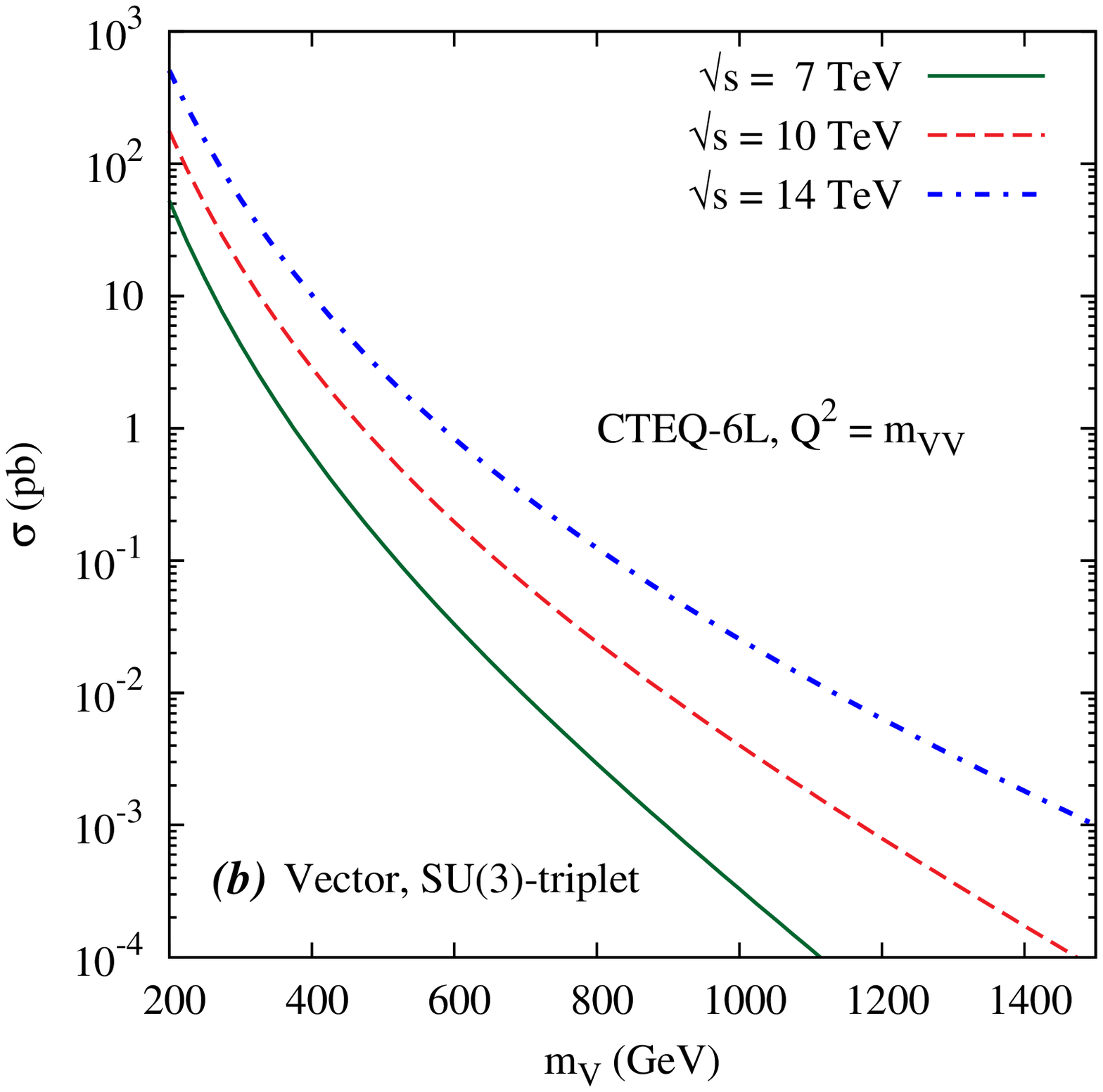}
\vspace*{-13ex}
\caption{\em Production cross-sections for scalar (left) and vector (right) 
         particles at the LHC for different collision energies.}
\label{fig:xsec}
\end{figure}
As is well-known, considerations of the $\rho$-parameter stipulates that the
mass-splitting between members of a $SU(2)_L$ must be small. Thus, all other 
quantum numbers being equal, the $SU(2)_L$ singlet diquark would have the 
smallest QCD production cross section. Similarly, for all coloured $SU(2)_L$ singlet 
scalars, the one in the $SU(3)_c$ triplet representation would have the lowest 
production cross-section. 

As for vector diquarks, various inequivalent choices for the kinetic
term (and, hence, for the coupling with the gluon) are possible (see
Ref.\cite{vector}).  Restricting ourselves to the case of minimal
coupling\footnote{Note that while this choice results in a smaller
  production cross section than is the case for a Yang-Mills type
  coupling, it is possible to get marginally smaller
  cross sections by tuning the anomalous
  dipole and quadrupole moments.}, 
we find that the resultant
cross-sections are significantly larger than is the case for the
corresponding scalar (see Fig.\ref{fig:xsec}). Thus, the choice of the
squark as our template diquark is the most conservative one.

\begin{figure}[tbp]
\vspace*{-5ex} \includegraphics[width=0.5\textwidth,
  height=0.5\textwidth]{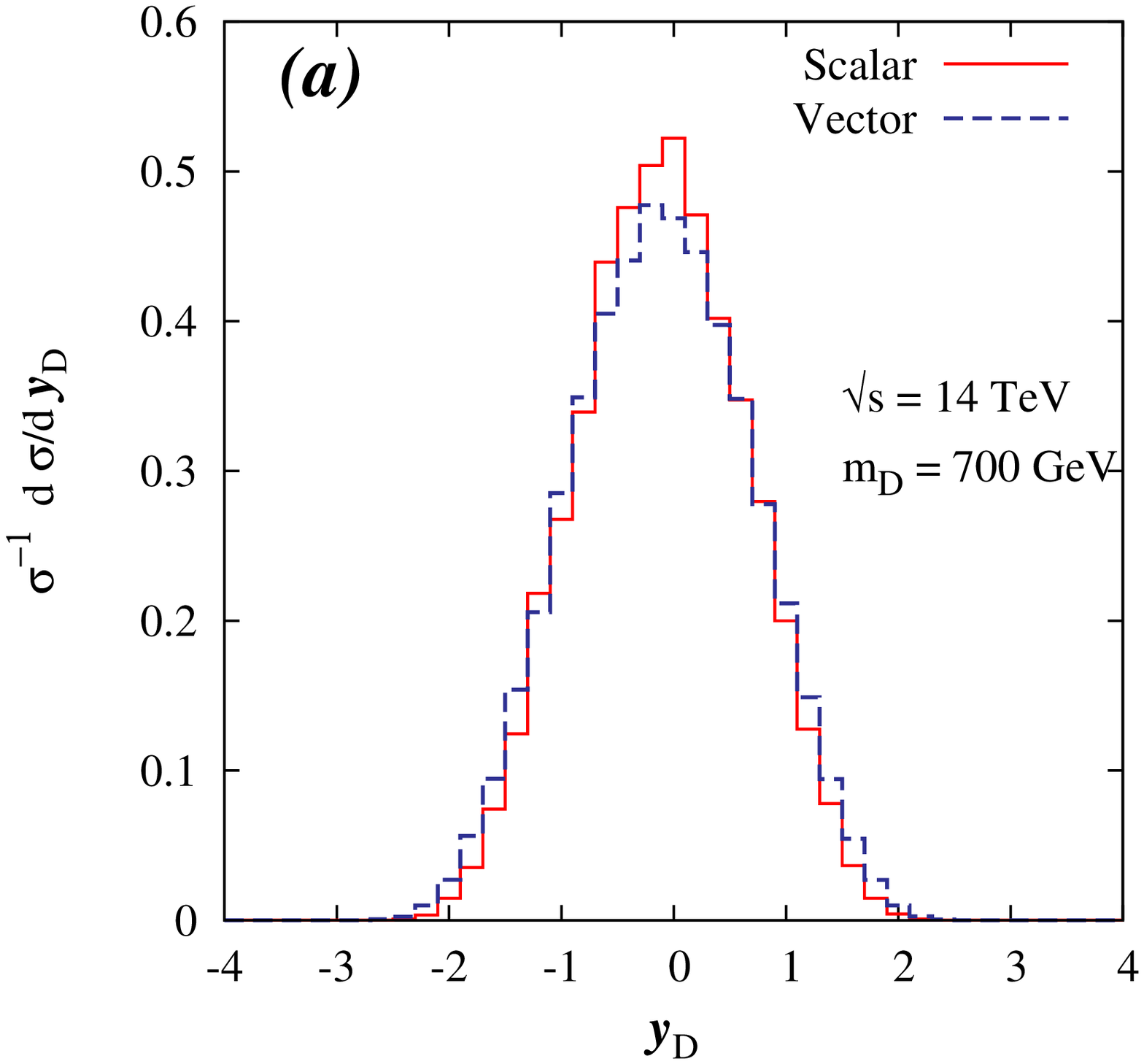} \hfill
\includegraphics[width=0.5\textwidth,
  height=0.5\textwidth]{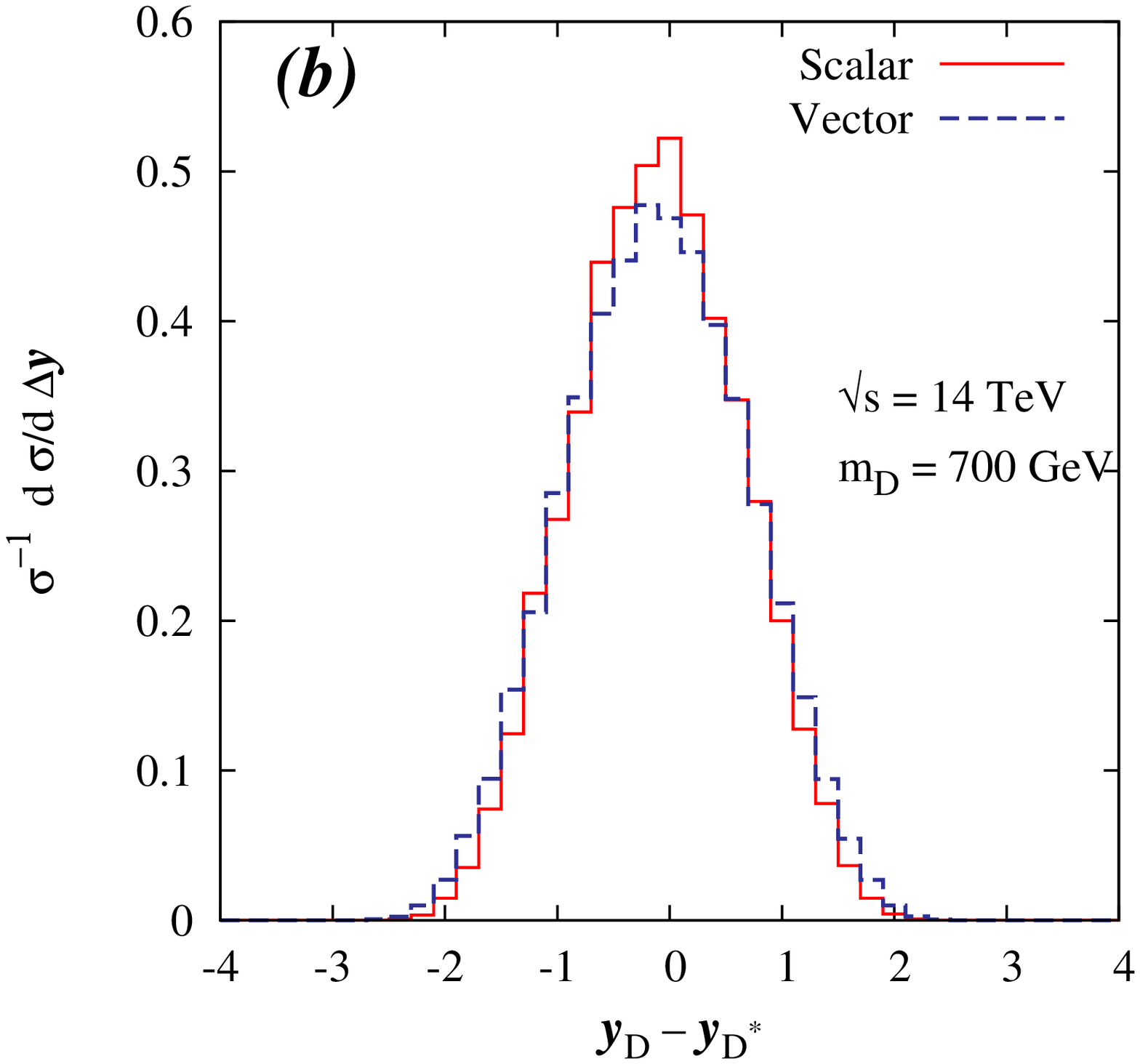}

\vspace*{-15ex}
\includegraphics[width=0.5\textwidth, height=0.5\textwidth]{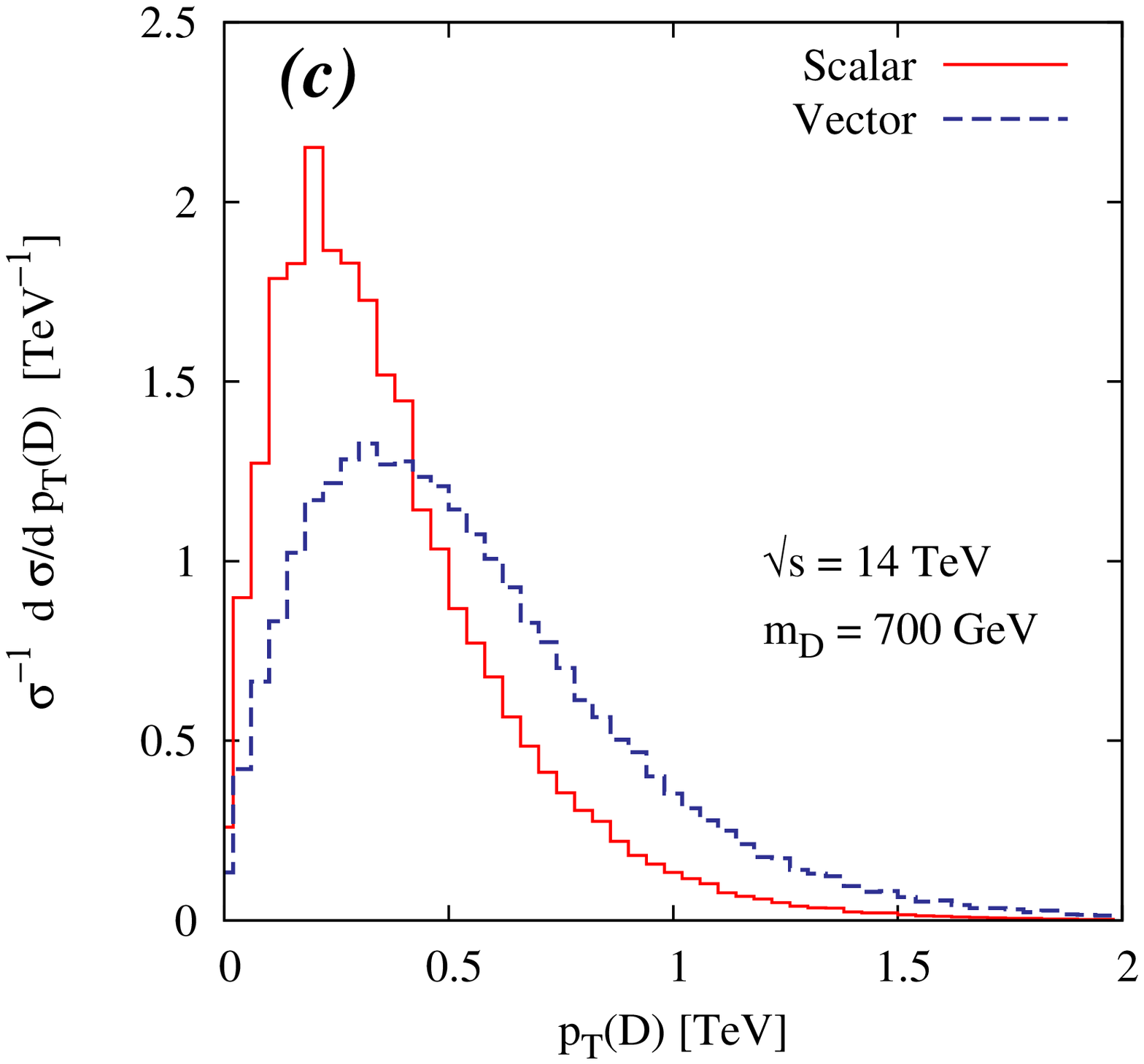}
\hfill
\includegraphics[width=0.5\textwidth, height=0.5\textwidth]{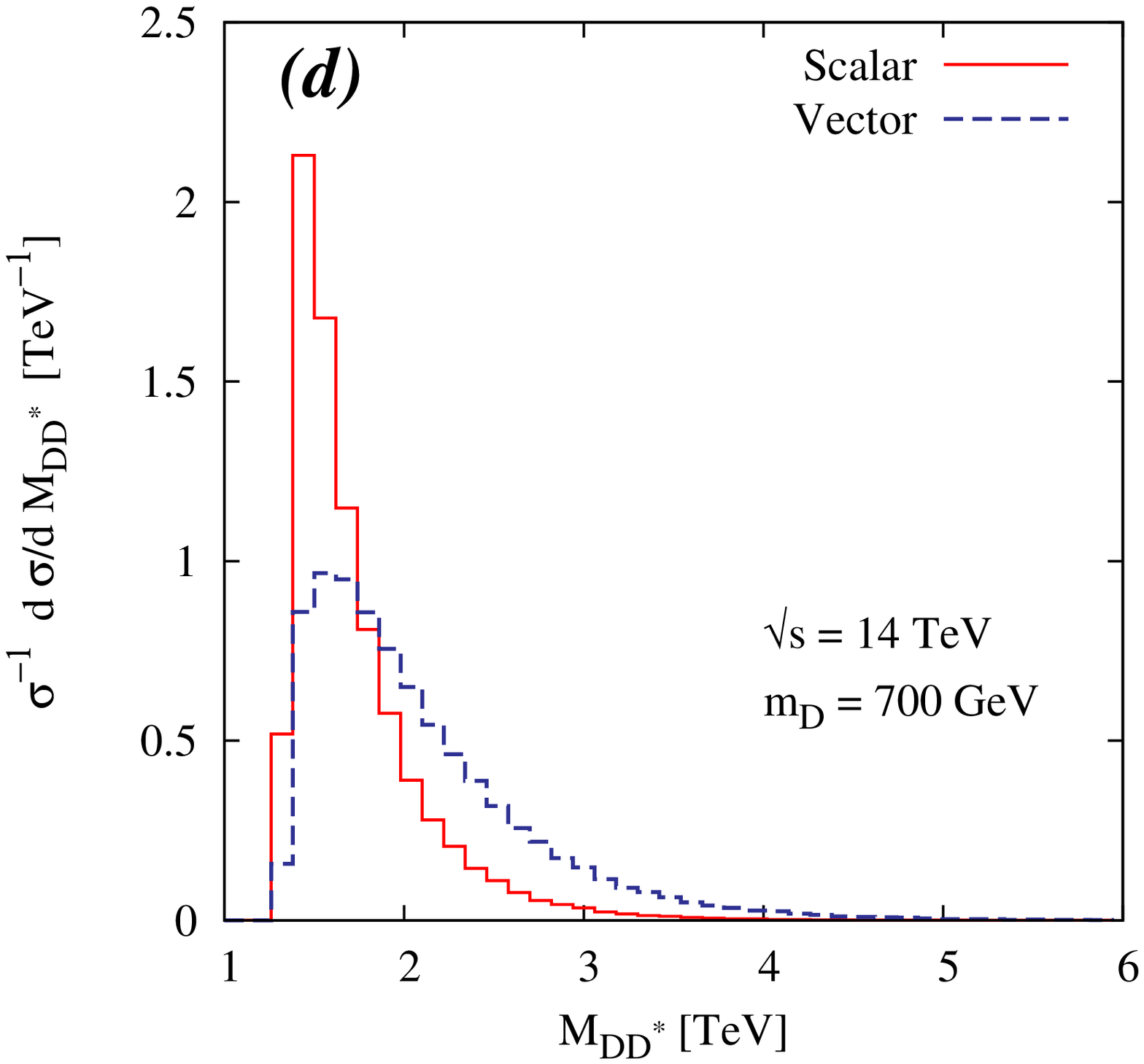}

\vspace*{-14ex}
\caption{\it Normalized kinematic distributions for QCD-driven pair-production 
of a diquark pair. $(a)$ the rapidity; $(b)$ the rapidity gap; 
$(c)$ the transverse momentum; and $(d)$ the invariant mass.}
\label{fig:distr}
\end{figure}

Of course, the signal size at a collider detector depends not only on
the total cross section, but on the kinematic distributions as
well. It is worthwhile to note---see Fig.\ref{fig:distr}$(a,b)$---that
the scalar and vector diquark have very similar angular
distributions. In other words, the rapidity-dependence of the
efficiencies would be very similar for the two cases.  On the other
hand, the spectrum is decidedly harder---see
Fig.\ref{fig:distr}$(c,d)$---for the vector case, a reflection of the
momentum dependence of its coupling to the gluon. Since we would need
to impose strong $p_T$-cuts, the overall efficiency would be better
for the vector case. Thus, on this count too, the choice of the squark
as a template is a conservative one.

Several searches---both in {\em R-parity} conserving and violating 
scenarios---have been made at LEP \cite{LEP_result} and Tevatron \cite{Tev_result}
without success. These searches claim to exclude, depending on the relevant 
SUSY parameters, $m_{\stop}$ upto $\sim 200 \gev$ assuming $100\%$ branching 
ratio for the respective decays. Such bounds, of course, weaken in the 
presence of other decay channels \cite{spd_ad_mm}. 

The high statistics data expected from the Large Hadron Collider (LHC) at 
$\surd{s} = 14 \tev$ provides an attractive option of looking for the $\stop$ in its 
$R$-parity violating decay mode(s). At the partonic level, our signal events would read
$$pp \rarw \stst\rarw (\bar{b}\bar{q})(bq) \ .$$
We use \textsc{Pythia6} \cite{pythia} to generate both
signal and background ( $\tt$, $\ww$, $ZZ$ ) events. 
Pure QCD background events ($\bb$+2-jets) have been 
generated at the parton level with \textsc{Alpgen2} \cite{alpgen} and 
interfaced with \textsc{Pythia}.

In \textsc{Pythia}, partons in an event go through the quarks from hard 
processes undergo parton shower development and hadronization 
followed by decays. \textsc{Pythia} also emulates the initial (ISR) and 
final state radiation (FSR) from the partons for all the event samples.
Tools provided by
\textsc{Pythia} have been used to define a toy calorimeter with the
broad features of the LHC detectors: pseudo-rapidity coverage 
($-4.5\le\eta\le 4.5$) and segmentation ($\Delta\eta\times\Delta\phi =
0.1\times 0.087$) \cite{CMS_ATLAS}. 
To simulate the finite energy resolution, the jet momenta are 
  smeared with a Gaussian function with a energy-dependent width $\delta E$
  given by
\[
    \frac{\delta E}{E} = \frac{1}{\sqrt{E \, (\gev)}} \oplus 0.05
\]
with the two contributions being added in quadrature. 
The final state particles are passed through this toy calorimeter for
forming jets. We also look for leptons ($\ell = e, \mu$) but within
the more restricted tracking coverage of the LHC detectors. Since
leptons are measured quite accurately, we use the generated values of
their energy, momentum and direction. Events are reconstructed in the
following manner: 
\bitem
\item	Calorimeter cells with $\et > 1\gev$ are used for jet 
        reconstruction. The cone algorithm with, 
	$\Delta R = \sqrt{(\Delta\eta)^{2}+(\Delta\phi)^{2}} = 0.7$, 
	has been used requiring $\etjet \ge 30 \gev$. 
\item	Leptons with $-2.4\le\eta\le 2.4$ and $\et\ge 5 \gev$ are deemed 
	identified.
\item 	A jet originating from a {\em b}-quark ({\em b}-jet) is identified 
        by the decay length of the B-hadron(s) within it. The efficiency for 
        tagging {\em b}-jets is tuned to the expected value ($\sim 50\%$ for 
	{\em b}-jets in $t\bar{t}$ events).
\eitem

\begin{figure}[!h]
\vspace*{-2ex}
\includegraphics[width=0.5\textwidth]{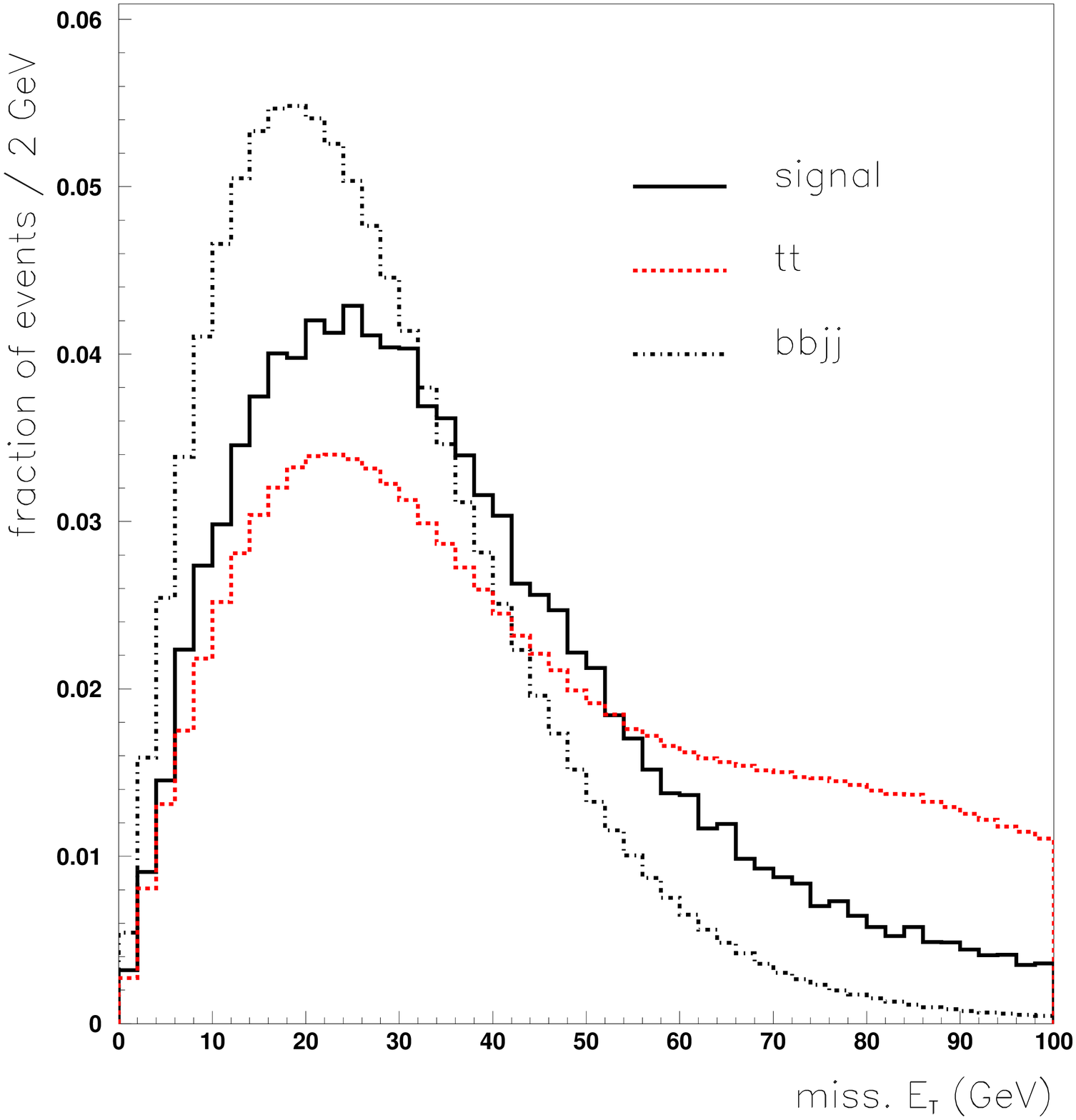}
\hfill
\includegraphics[width=0.5\textwidth]{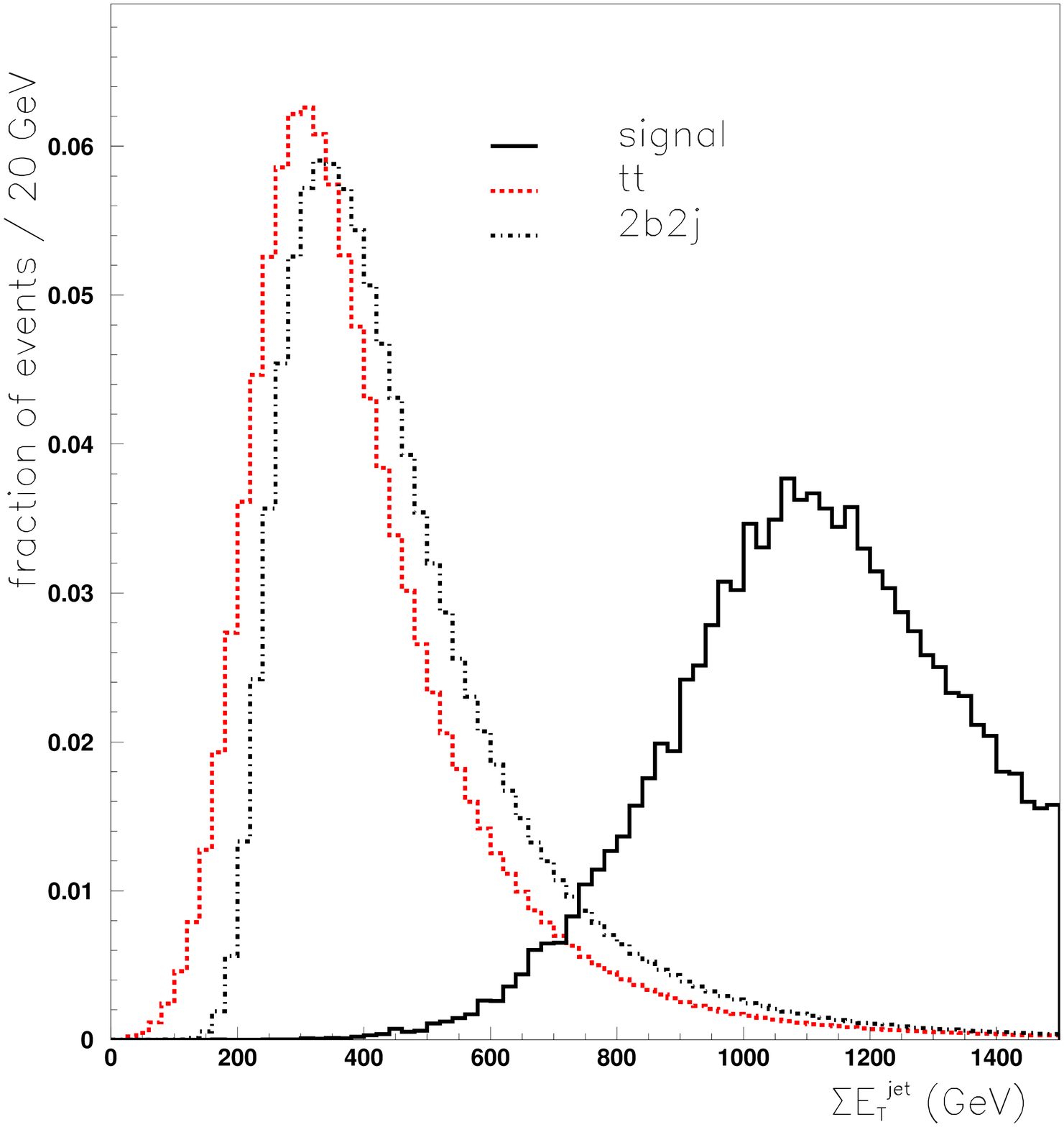}
\vspace*{-6ex}
\caption{\it Distribution of $\met$ and $\sum E_{T}^{jet}$ for signal events
	 ( $m_{\stop} = 550 \gev$ ) and two major background sources before 
	 any selection has been applied.}
\label{fig:selcuta}
\end{figure}
For the signal events, we expect four jets with high $\et$ and no
lepton with a high $\pt$. We should not expect any $\met$ except for
that arising due to mis-measurement of the jets and neutrinos from
secondary and tertiary decays of long lived particles. There may be
additional objects (extra jets, $\met$, leptons, etc) arising from 
underlying events (a serious concern at the high luminosity runs of the LHC) 
as well as due to ISR and FSR. But such objects are
expected to fail hard cuts used in this analysis.  At this juncture,
we could have considered imposing an upper cut on $\met$ to remove,
say, the $t \bar t$ background. However, since the jets from the
$\stop$ decays have a high $\et$, even a small relative
mis-measurement could result in a substantial $\met$ leading to the
rejection of a large fraction of the signal events (see
Fig.\ref{fig:selcuta}).  Given this, and the relative unimportance of
the $t \bar t$ background, we choose not to impose such a criterion.

\begin{figure}[tbp]
\vspace*{-4ex}
\includegraphics[width=0.5\textwidth]{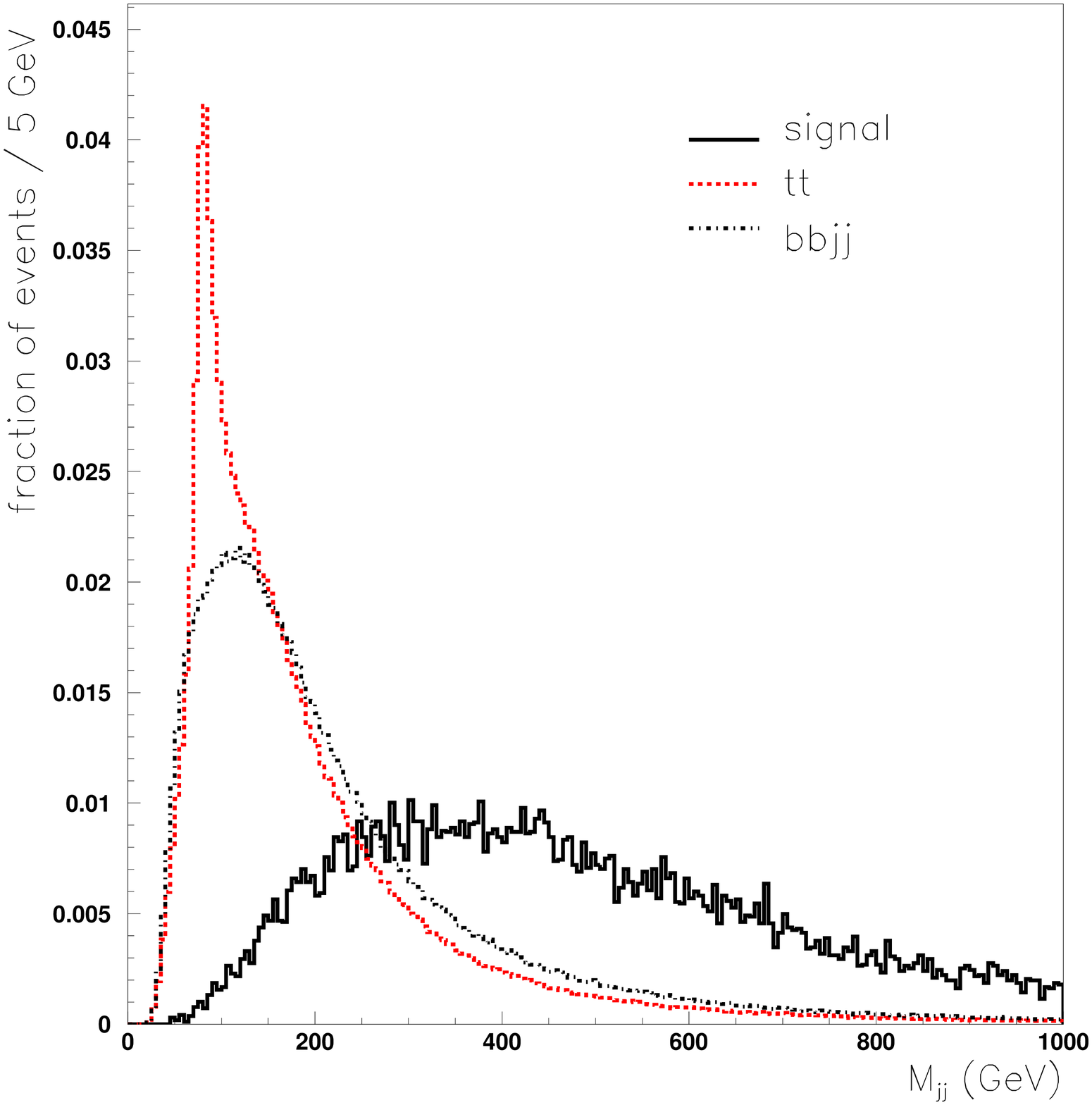}
\includegraphics[width=0.5\textwidth]{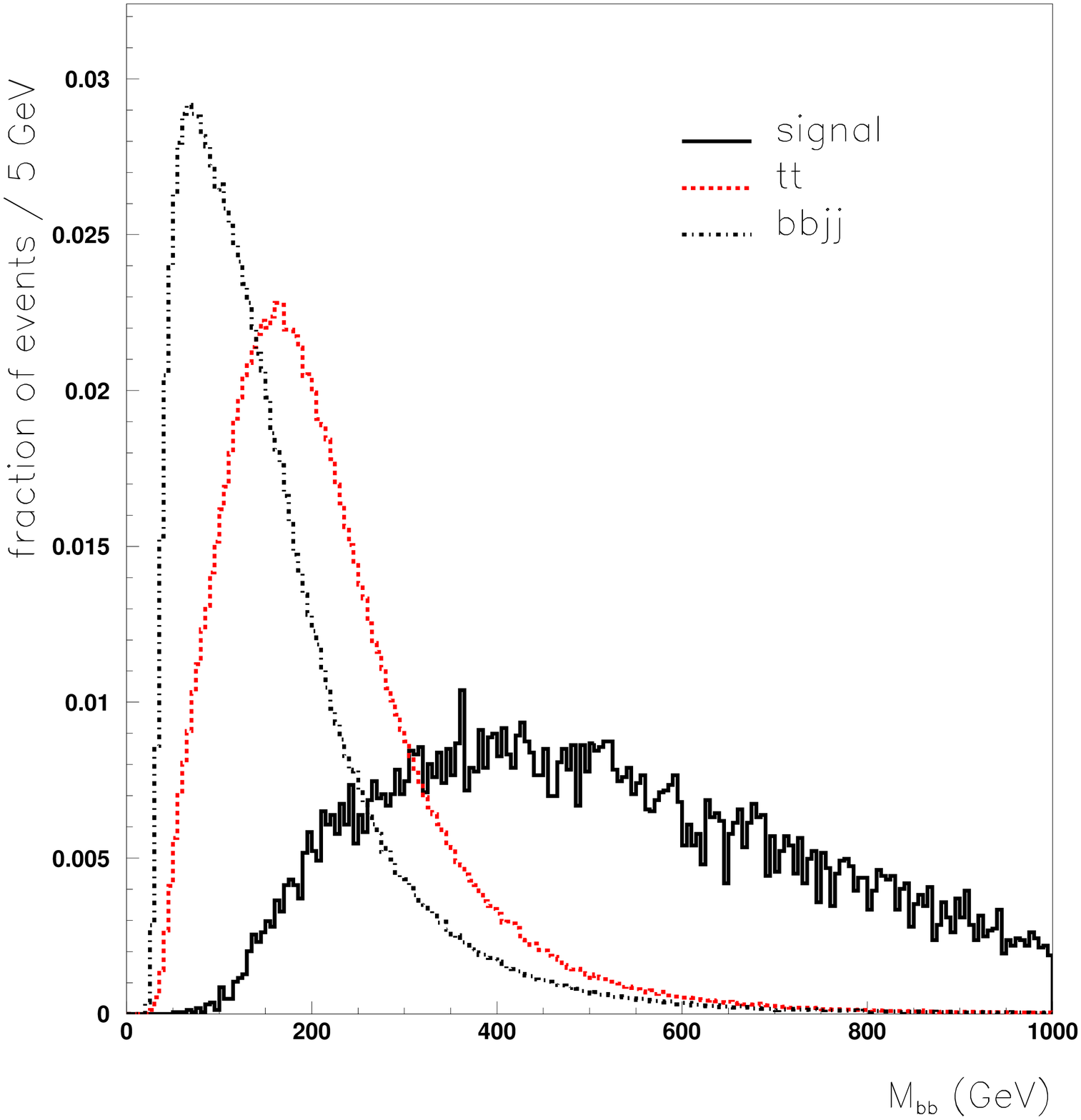}
\vspace*{-6ex}
\caption{\it Distribution of $M_{jj}$ and $M_{bb}$ for signal events 
         ($m_{\stop} = 550 \gev$) and two major background sources after the 
         cut on the number of ordinary and b-tagged jets has been applied.}
\label{fig:selcutb}
\end{figure}

Instead, we use the following criteria for selecting signal events and rejecting 
backgrounds.
\bnum
\item 	There should be only four jets in the event ($N_{jet} = 4$)
        with 
	$\etjet \ge E_{T\, {\rm min}}^{jet}$. Of these, two are tagged b-jets 
	($b_{1}, ~b_{2}$) while the other two ($j_{1}, ~j_{2}$) are ordinary jets, 
	i.e., they do not satisfy b-tagging criteria. ({\bf Sel 1}). \\
        The jets are ordered descending in $\et$.
\item	Events containing any isolated lepton with $\pt \ge 15 \gev$ are rejected.
	({\bf Sel 2})
\item	We also require 
        $m_{b_{1}b_{2}}, m_{j_{1}j_{2}} \not\in \{70, 100\}\gev$ 
	( see plots in 
  	Fig.\ref{fig:selcutb} ) to reject events with 
	genuine $W$ or $Z$ ({\bf Sel 3}). The plot for $M_{jj}$ for $t\bar{t}$ 
	events in 
	Fig. \ref{fig:selcutb} expectedly shows the $W$ mass peak.
\item 	To eliminate the huge soft QCD background which may mimic signal with 
	fake $b$-jet tagging, we require (see right plots in Fig.
	\ref{fig:selcuta}) 
	$\sum \etjet \ge S_{\rm min}$. ({\bf Sel 4})
\enum

For each value of $\mstop$, nine different combinations of {\bf Sel 1}
and {\bf Sel 4}, {\em viz}, $E_{T\, {\rm min}}^{jet} = $ \{30, 50,
75\} GeV and $S_{\rm min} = $ \{400, 500, 600\} GeV along with {\bf
  Sel 2} and {\bf Sel 3}, were tried and the combination that gave the
best signal visibility has been used for the final result.  It may be
noted that, in this analysis, we have used calorimetric observables
which may be estimated fairly well both in magnitude and direction
using parametric simulation.

\begin{figure}[tbp]
\vspace*{-4ex}
\includegraphics[width=0.45\textwidth]{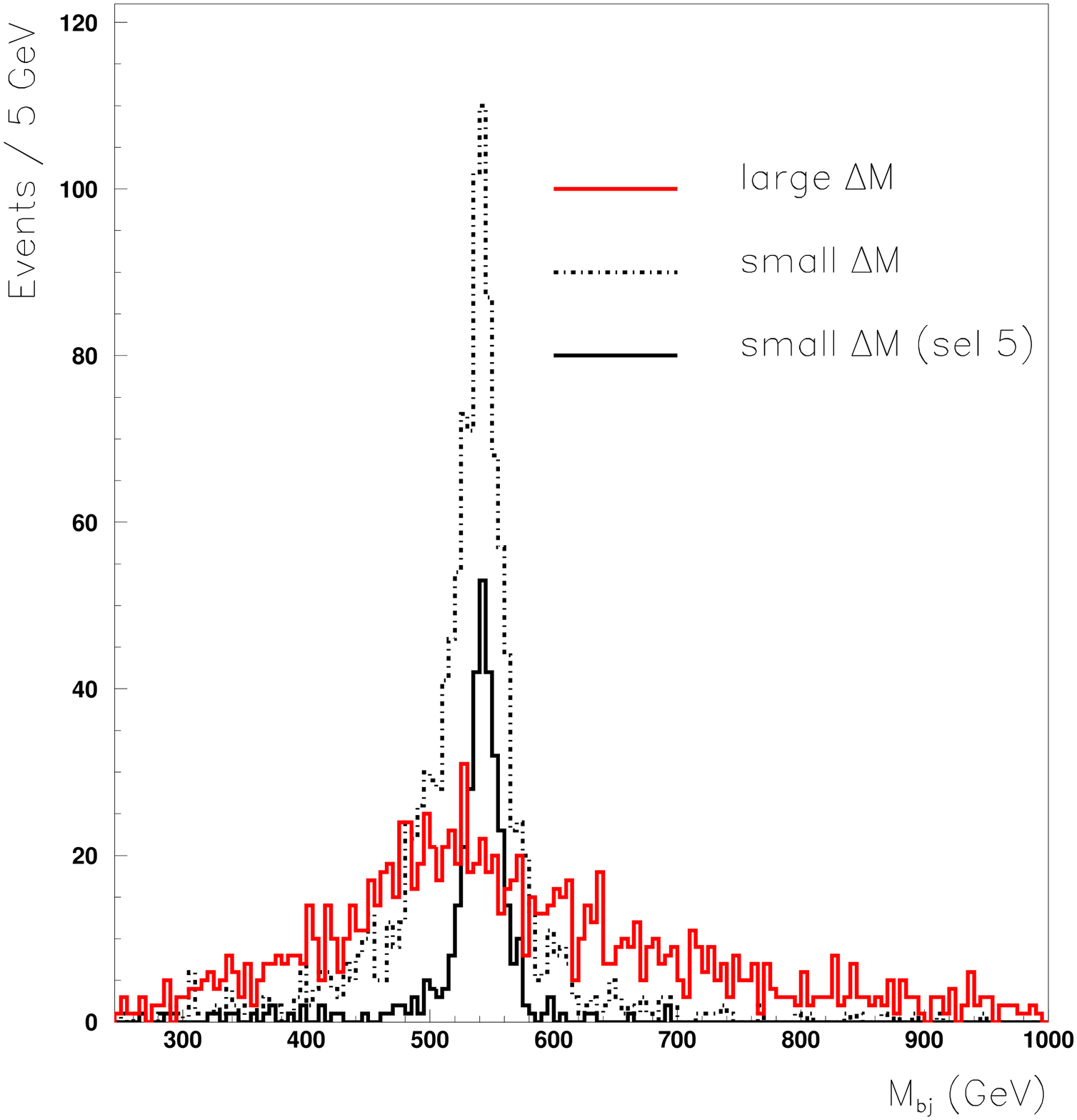}
\hfill
\includegraphics[width=0.45\textwidth]{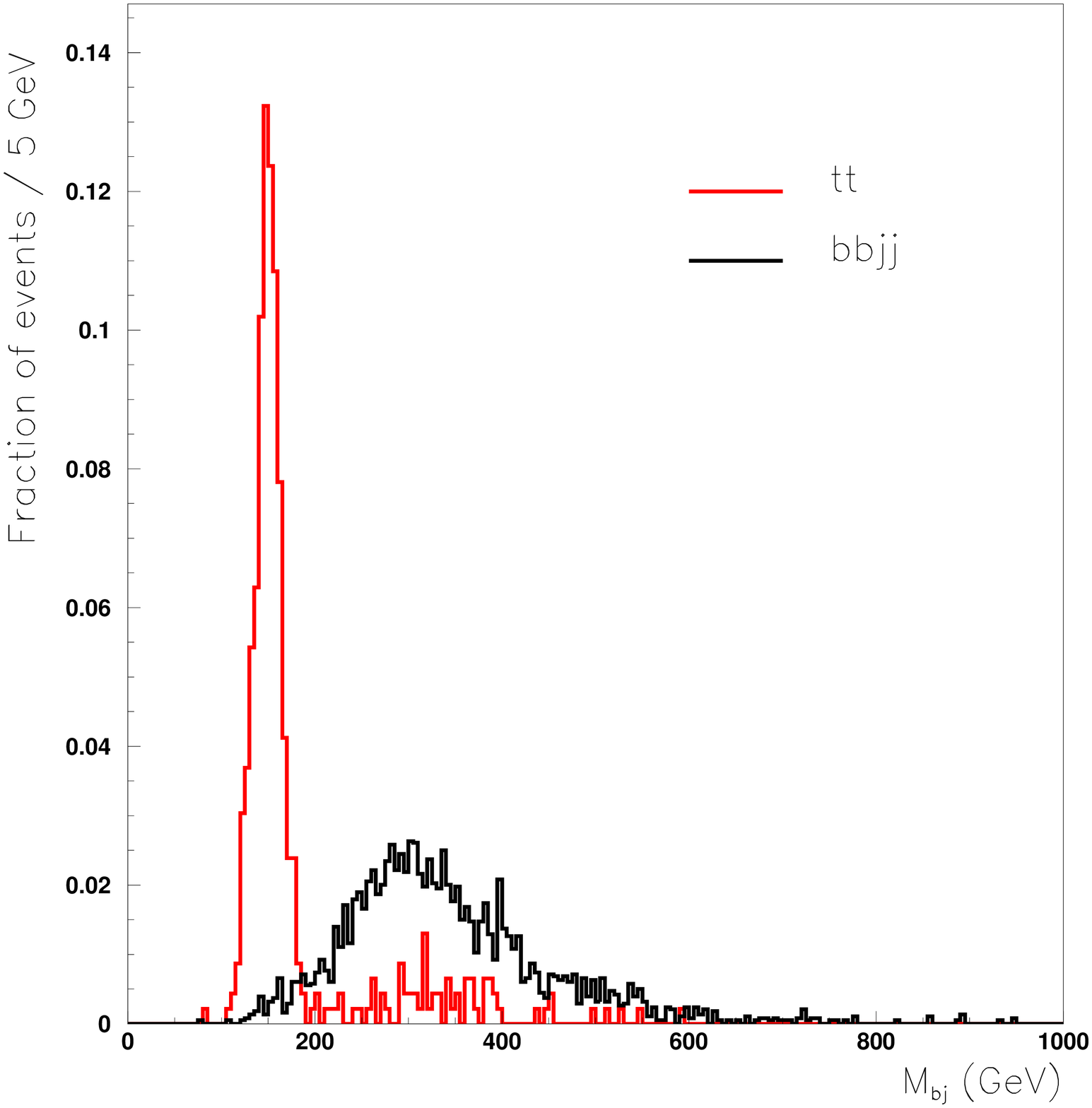}
\vspace*{-2ex}
\caption{\it The left plot shows distributions ( normalized to arbitrary 
	 luminosity ) of $m_{right}^{avg}$ (black dashed line) and 
         $m_{wrong}^{avg}$ (red solid line ) for signal events 
         ($m_{\stop} = 550 \gev$) along with $m_{right}^{avg}$ after 
         Sel 5 (black solid line). The right plot shows distributions 
         of $m_{right}^{avg}$ (normalized to shape) after Sel 5 for two 
         major background sources.}
\label{fig:stopmass}
\end{figure}

The events are sought to be reconstructed by computing the invariant
masses of a pair of jets, one being tagged a $B$-jet, and the other
not so. Clearly, four such pairings occur, viz.  $\{ b_{1}j_{1},
b_{2}j_{2}, b_{1}j_{2}, b_{2}j_{1}\}$.  Either $b_{1}j_{1},
b_{2}j_{2}$ ({\bf combination 1}) or, $b_{1}j_{2}, b_{2}j_{1}$ ({\bf
  combination 2}) represents the {\em right combination} in terms of
the quarks from the hard processes (the decays) while the other
represents the {\em wrong combination}. The two invariant masses
belonging to the right combination would be expected to have a
difference smaller than that for the wrong combination. To further
reduce the chance of choosing the wrong combination due to accidental
large mismeasurement of jet energies, we require ({\bf Sel 5})
$$\Delta m = min\{|m_{b_{1}j_{1}} - m_{b_{2}j_{2}}|, 
                 |m_{b_{1}j_{2}} - m_{b_{2}j_{1}}|\} \le 20\gev \ .$$ 
The average mass of the right combination $m_{right}^{avg}$: 
$(m_{b_{1}j_{1}} + m_{b_{2}j_{2}})/2$ or
$(m_{b_{1}j_{2}} + m_{b_{2}j_{1}})/2$ as specified by Sel 5 is then plotted 
in Fig.\ref{fig:stopmass}. The other average is named $m_{wrong}^{avg}$. 
The plot for $m_{right}^{avg}$ in signal events shows the peak close to the 
input value of $m_{\stop}$ while the plot for $m_{wrong}^{avg}$ 
(the wrong combination) is expectedly much wider though it may contain some 
`right' combinations due to accidental large fluctuation.

Finally, we proceed to estimate the signal significance ({\em S}) using signal 
and background events passing all the selection cuts mentioned above and use 
leading order cross-sections for signal as well as background processes.
The effective cross-section for the signal events is
$$\sigma_{eff} = 
         \sigma(pp\rarw\stop \, \stop^{*}) B^{2} \ , $$
where $B$ is the total branching 
ratio\footnote{Note that limits from flavour changing neutral current
    processes constrain products of $\lambda''$
    couplings~\cite{bcs_debrupa}. The present study, though, is
    insensitive to how many channels share $B$.  } 
for the $R$-parity violating decays of the $\stop$ involving
a $\bar{b}$-quark, namely $\stop \rarw \bar{b}\bar{d},
\bar{b}\bar{s}$. It may be noted that the fraction of signal events
passing the selection criteria increases with $\mstop$ although
$\sigma(pp\rarw\stst)$ falls rapidly.  We choose a $25
\gev$ wide window in the $m_{right}^{avg}$ distribution in which the
signal is most abundant and has the right shape and determine the
fraction of generated signal events ($f_{sig}$) falling within this
window.  Finally the estimated number of signal events is obtained
through \beqn N_{S} = f_{sig}.\sigma_{eff}.\lint \label{eqn:signal}
\ , \eeqn where $\lint$ is the assumed value of integrated luminosity.
Similarly, for each background process the number of events falling
within the invariant mass window is determined and added up to obtain
$N_{B}$.  The statistical significance may be estimated as \beqn S =
\frac{N_{S}}{\sqrt{N_{B}}} \ .  \label{eqn:signi} \eeqn

\begin{figure}[!t]
\bcent
\includegraphics[width=0.4\textwidth, angle=-90]{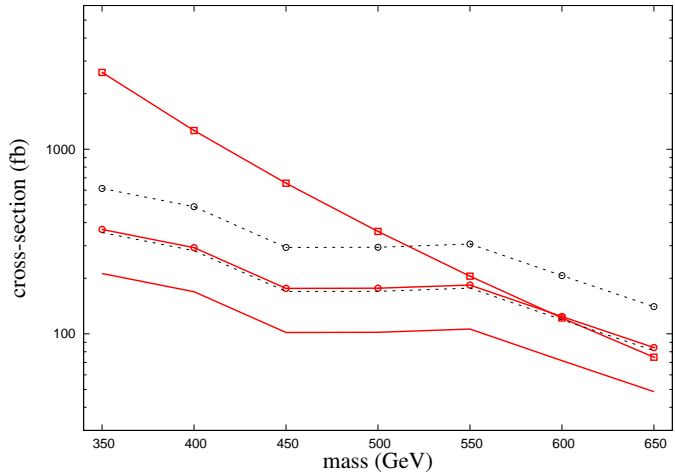}
\ecent
\caption{\it The 
	 minimum value of $\sigma_{eff}$ (see text) required 
	 for excluding signal at $3\sigma$ for 
	 $\lint = 100 ~fb^{-1}$ (solid line with open circles) and 
         $\lint = 300 ~fb^{-1}$ (solid line). The same required for 
	 observing a $5\sigma$ excess over the background are shown for
	 $\lint = 100 ~fb^{-1}$ (dashed line with open circles) and 
         $\lint = 300 ~fb^{-1}$ (dashed line). For comparison,
         the solid line with open squares shows 
	 $\sigma(pp\rarw\stst)$.}
\label{fig:xsecbr}
\end{figure}

Next, we use eqns.(\ref{eqn:signal} \& \ref{eqn:signi}) to
calculate the minimum value of effective signal cross-section
($\sigma_{eff}^{min}$) required to obtain $S = 3$ and $S = 5$ for a
given luminosity (we use $\lint = 100\fb^{-1}, \quad
300\fb^{-1}$). The values of $\stst$ production cross-section and
$\sigma_{eff}^{min}$ have been listed in Table \ref{tab:xsecbr} and plotted
in Fig. \ref{fig:xsecbr}. It is evident that a large range of
parameter space may be explored with $\lint = 100 \fb^{-1}$.

\begin{table}[!h]
\begin{tabular}{lccccc} 
\hline\hline
        & $\sigma(pp\rarw\stst)$ 
        & \multicolumn{4}{c}{$\sigma_{eff}^{min}$ (fb) required for} \\
\cline{3-6}
$m_{\stop}$ 
        &  (fb)
        & \multicolumn{2}{c}{S = 3} 
        & \multicolumn{2}{c}{S = 5}\\
    &   & $\lint = 100 ~fb^{-1}$ & $\lint = 300 ~fb^{-1}$
        & $\lint = 100 ~fb^{-1}$ & $\lint = 300 ~fb^{-1}$ \\
\hline
300 & 5876 &  494 & 285 & 824 & 476 \\
350 & 2606 &  368 & 212 & 613 & 354 \\
400 & 1262 &  293 & 169 & 488 & 282 \\
450 &  655 &  176 & 102 & 294 & 167 \\
500 &  359 &  177 & 102 & 294 & 170 \\
550 &  205 &  184 & 106 & 306 & 177 \\
600 &  122 &  124 &  72 & 207 & 119 \\
650 &   75 &   84 &  49 & 140 &  81 \\
\hline\hline
\end{tabular}
\caption{\it 
	Table shows the minimum value of $\sigma_{eff}$ (see text) required 
	for different values of $m_{\stop}$ (column 1) for exploring signal at 
	$3\sigma$ for $\lint = 100 ~fb^{-1}$ (column 3) and $\lint = 300 ~fb^{-1}$ (column 4). 
	The same required for observing a $5\sigma$ excess over the background are shown for
	$\lint = 100 ~fb^{-1}$ (column 5) and $\lint = 300 ~fb^{-1}$ (column 6). Production 
	cross-sections for the signal are shown in column 2.} 
\label{tab:xsecbr}
\end{table}
Jet reconstruction and determination of invariant mass of a particle
decaying into two or more jets become difficult at the LHC due to the
presence of underlying events, hard radiation from the partons and
large $\pt$ of the decaying particle. It has been claimed that a
better invariant mass reconstruction is possible using more
sophisticated jet reconstruction and analysis techniques and hence
increase the signal significance \cite{sub_jet}. As such techniques
mature, they will help enlarge the scope of this analysis.

In conclusion, we have explored the possibility of observing, at 
the LHC, a particle decaying into a pair of jets, of which one 
has a $b$-quark as the originator. Concentrating on 
the scalar top quark in a supersymmetric extension of the 
Standard Model, we demonstrate that simple and robust detector level 
observables may be used to explore a fairly wide range of the 
stop mass with integrated luminosities expected from the LHC. 
In addition, a fairly good estimate of the mass can also be 
obtained. And while our analysis has concentrated on a particular 
scenario, we have also demonstrated that it represents the most 
conservative choice amongst a class of such particles that appear 
naturally in a large variety of theories. Consequently, this 
analysis can be easily extended to such other scenarios 
allowing one to explore masses significantly larger than 
what this particular model allows us to. 
Aiding this assertion is the observation that, in many of such models
such as those incorporating diquarks or excited quarks, the branching 
fraction into the relevant decay mode is overwhelmingly the dominant one.

{\bf Acknowledgements:}
DC thanks the IPMU for hospitality as this work was
being finalized. This work was supported by the 
DST, India under project SR/MF/PS-03/2009-VB-I 
and the World
Premier International Research Center
Initiative (WPI Initiative), MEXT, Japan. 


\end{document}